\documentclass[10pt,letterpaper]{article}
\usepackage{fullpage}
\usepackage{amsmath,amssymb}
\usepackage{cite}
\usepackage{nameref,hyperref}
\usepackage{graphicx}

\usepackage[aboveskip=1pt,labelfont=bf,labelsep=period,justification=raggedright,singlelinecheck=off]{caption}

\makeatletter
\renewcommand{\@biblabel}[1]{\quad#1.}
\makeatother

\date{}

\begin{document}

\vspace*{0.2in}

\begin{flushleft}
{\Large
\textbf\newline{Gain control with A-type potassium current: $I_A$ as a switch between divisive and subtractive inhibition} 
% Please use "sentence case" for title and headings (capitalize only the first word in a title (or heading), the first word in a subtitle (or subheading), and any proper nouns).
}
\newline
\\
Joshua H Goldwyn\textsuperscript{1*},
Bradley R Slabe\textsuperscript{2},
Joseph B Travers\textsuperscript{3},
David Terman\textsuperscript{2}
\\
\bigskip
\textbf{1} Department of Mathematics and Statistics, Swarthmore College, Swarthmore, Pennsylvania, USA
\\
\textbf{2} Department of Mathematics, The Ohio State University, Columbus, Ohio, USA 
\\
\textbf{3} Division of Biosciences, College of Dentistry, The Ohio State University, Columbus, Ohio, USA
\\
\bigskip

% Use the asterisk to denote corresponding authorship and provide email address in note below.
* Corresponding author: jhgoldwyn@gmail.com

\end{flushleft}

\section*{Abstract}

Neurons process and convey information by transforming barrages of synaptic inputs into spiking activity.  Synaptic inhibition typically suppresses the output firing activity of a neuron, and is commonly classified as having a {\it subtractive} or {\it divisive} effect on a neuron's output firing activity.  Subtractive inhibition can narrow the range of inputs that evoke spiking activity by eliminating responses to non-preferred inputs.  Divisive inhibition is a form of gain control: it modifies firing rates while preserving the range of inputs that evoke firing activity.  Since these two ``modes'' of inhibition have distinct impacts on neural coding, it is important to understand the biophysical mechanisms that distinguish these response profiles.

In this study, we use simulations and mathematical analysis of a neuron model to find the specific conditions (parameter sets) for which inhibitory inputs have subtractive or divisive effects.  Significantly, we identify a novel role for the A-type Potassium current ($I_A$).  In our model, this fast-activating, slowly-inactivating outward current acts as a switch between subtractive and divisive inhibition.  In particular, if $I_A$ is strong (large maximal conductance) and fast (activates on a time-scale similar to spike initiation), then inhibition has a subtractive effect on neural firing.  In contrast, if $I_A$ is weak or insufficiently fast-activating, then inhibition has a divisive effect on neural firing.  We explain these findings using dynamical systems methods (plane analysis and fast-slow dissection) to define how a spike threshold condition depends on synaptic inputs and $I_A$.

Our findings suggest that neurons can ``self-regulate'' the gain control effects of inhibition via combinations of synaptic plasticity and/or modulation of the conductance and kinetics of A-type Potassium channels.  This novel role for $I_A$ would add flexibility to neurons and networks, and may relate to recent observations of divisive inhibitory effects on neurons in the nucleus of the solitary tract.

\section*{Author summary}
Neurons process information by generating spikes in response to two types of synaptic inputs.  Excitatory inputs increase spike rates and inhibitory inputs decrease spike rates (typically).  The interaction between these two input types and the transformation of these inputs into spike outputs is not, however, a simple matter of addition and subtraction.  Inhibitory inputs can suppress outputs in a variety of ways.  For instance, in some cases, inhibition adjusts the rate of spiking activity while preserving the range of inputs that evoke spiking activity; an important computational principle known as gain control.  We use simulations and mathematical analysis of a neuron model to identify properties of a neuron that determine how inhibitory inputs affect spiking activity.  Specifically, we demonstrate how the gain control effects of inhibition depend on the A-type Potassium current.  This novel role for the A-type Potassium current provides a way for neurons to flexibly regulate how they process synaptic inputs and transmit signals to other areas of the brain.

% Use "Eq" instead of "Equation" for equation citations.
\section*{Introduction}

The activity of a neuron is driven by barrages of synaptic inputs.
Synaptic inputs are classified as either excitatory (those that promote spike generation) and inhibitory (those that impede spike generation).
The interplay between these two ``opposing'' inputs influences how neurons process and transmit information in the brain.

\begin{figure}[!h]
\centering
\includegraphics[width=\textwidth]{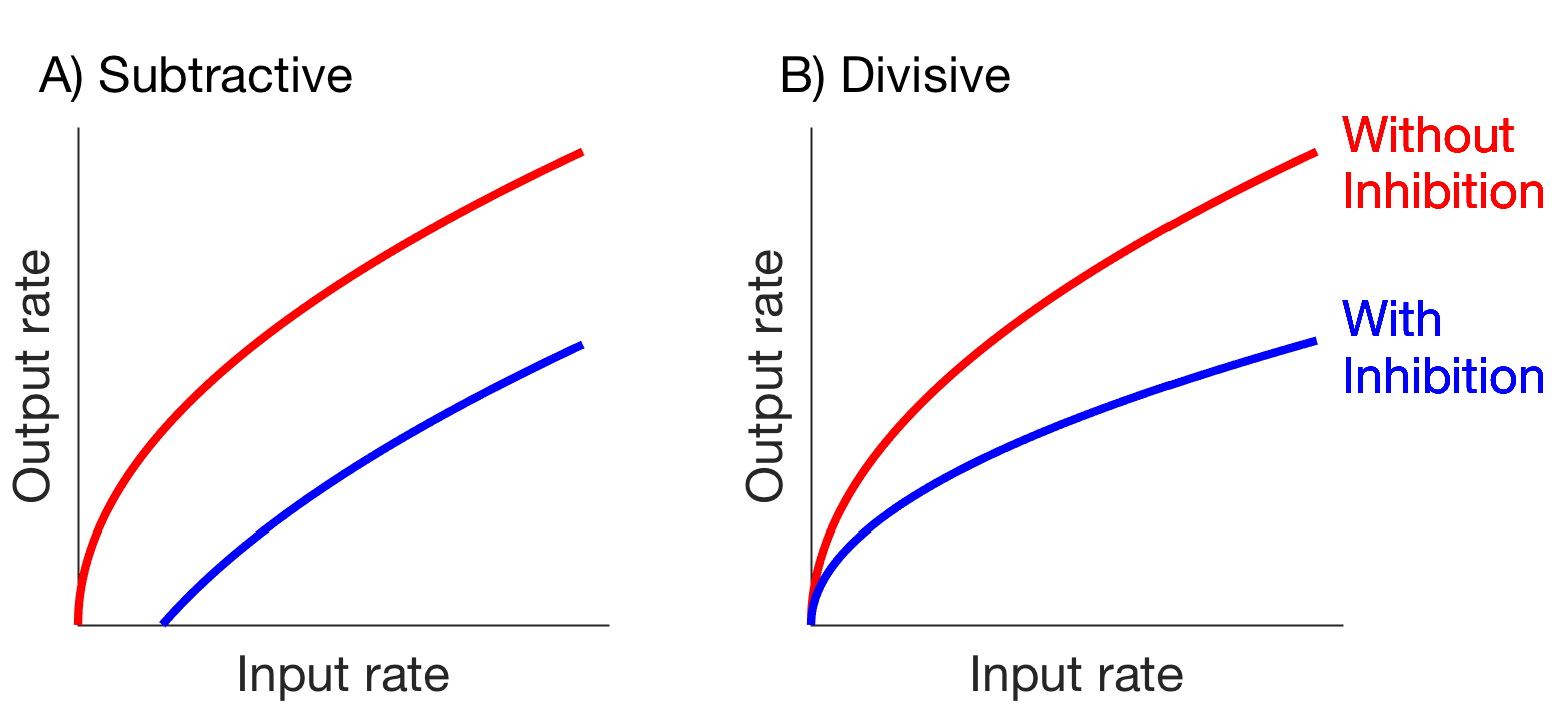}
\caption{Comparison of firing rate input/output relations for subtractive and divisive inhibition (illustration only, not actual data).  {\bf A:} Subtractive inhibition: output rate without inhibition is $r_{out} = \sqrt{r_{in}}$, and output rate with inhibition is $r_{out}  = \sqrt{r_{in}}-c$, where $c$ is a constant with $c>0$.
{\bf B: } Divisive inhibition: output rate without inhibition is $r_{out} = \sqrt{r_{in}}$ (same as in {\bf A}), and output rate with with inhibition is $\alpha \sqrt{r_{in}}$, where $\alpha$ is a constant with $0<\alpha < 1$.}
 \label{fig:intro}
\end{figure}

To characterize the nature of inhibition, researchers often distinguish between inhibition that has a \emph{subtractive} effect on neural firing, versus inhibition that has a \emph{divisive} effect~\cite[for review]{Silver2010}.
Inhibition is said to be subtractive if it reduces the firing activity of a neuron by (roughly) a constant amount, regardless of the strength or amount of synaptic excitation.
Inhibition is said to be divisive if it reduces the firing activity of a neuron by an amount that is (roughly) proportional to the neuron's firing rate.  
We illustrate this distinction in Fig~\ref{fig:intro}, by showing output firing rate of a neuron as a function of the rate of its excitatory inputs (not actual data).

% Place figure captions after the first paragraph in which they are cited.

The differences between these modes of inhibition has important consequences for neural coding.  
Subtractive inhibition suppresses responses to ``non-preferred'' stimuli that evoke infrequent responses in the absence of inhibition.  This can be useful for promoting the representation of ``preferred'' inputs.  
In contrast, divisive inhibition is a mechanism for neural gain control:  it reduces the firing rate of a neuron while preserving the overall range of inputs to which the neuron is responsive~\cite{Wilson2012}.
Understanding the physiological mechanisms that determine how and why inhibition acts in these two modes is key for understanding how neurons and networks function.
Past studies have identified numerous possibilities for mechanisms underlying these two modes of inhibition, including the stochastic (noisy) nature of synaptic inputs~\cite{Doiron2000}, the balance between excitatory and inhibitory inputs~\cite{Chance2002}, shunting inhibition~\cite{Prescott2003, Mitchell2003}, synaptic depression~\cite{Rothman2009}, and circuit structure~\cite{Wilson2012, Mejias2014}, and see~\cite{Silver2010} for additional review.  

In this study, we identify a novel, neuron-level mechanism that affects whether inhibition acts in a subtractive or divisive manner. 
Through mathematical analysis and simulations, we explore the combined effects of synaptic inputs and voltage-gated ion currents on spiking dynamics of a neuron model.
We find that A-type potassium current can control whether inhibition has a divisive or subtractive effect on firing rate activity. 
This current is a source of voltage-gated, dynamic, negative feedback. 
If it is sufficiently large and activates rapidly, then it combines with inhibitory inputs to suppress firing activity in a subtractive manner.  
If, instead, the A-current is sufficiently weak or activates slowly (relative to spike initiation dynamics), then inhibition has a divisive effect on firing rates.  

Our work identifies a route through which adaptive or dynamic changes to the intrinsic dynamics of neurons (for example, through modification of ion currents~\cite{Levitan1988}) can modulate the effects of inhibition.  This capability for individual neurons to switch between different inhibition ``regimes'' could provide useful flexibility to neural systems.

\section*{Materials and methods}

We simulate and analyze two models of neural dynamics.  The first is a one-compartment model that approximates a neuron as a single, isopotential unit (a ``point neuron'' model).  The second is a multi-compartment model that includes a region of voltage-gated currents attached to a spatially-extended region of passive membrane (``soma'' and ``dendrite'' regions, respectively).   We describe these models below.

\subsection*{One-compartment neuron model}

The dynamics of membrane potential, $V$, in the one-compartment neuron model are

\begin{align}
\label{eq:OneCptODE}
C V' &= - I_L - I_K - I_A - I_{Na} -  I_{Syn,E} - I_{Syn,I} 
\end{align}
where the membrane capacitance is $C=1~\mu\mbox{F/cm}^2$.
The ionic currents  (leak, potassium, A-type potassium, and sodium) are given by the equations

\begin{align}
I_L &= g_L (V-V_L), ~~~~~~~~~~ I_K = g_K n^4 (V-V_K), \nonumber \\
I_A &= g_A a^3 b (V-V_K), ~~~~~ I_{Na} = g_{Na} m^3 h (V-V_{Na}). 
\end{align}
We use the following fixed parameter values for maximal conductances:  $g_L = 1~\mbox{mS/cm}^2$, $g_K = 45~\mbox{mS/cm}^2$, and $g_{Na}= 37~\mbox{mS/cm}^2$.
We use a range of values for the maximal conductance of the A-current ($g_A$) to observe transitions between subtractive and divisive effects of inhibition.
The reversal potentials are $V_L = -70~\mbox{mV}$, $V_K = -80~\mbox{mV}$, and $V_{Na} = 55~\mbox{mV}$.

We make several simplifications, similar to those first suggested in \cite{Rinzel1978}, to the gating variables in the model.  
We identify sodium activation as a fast process and assume it evolves instantaneously to its voltage-dependent steady-state value. That is, we let  $m=m_\infty(V) = 1/(1 + e^{-(V+30)/15})$.  In addition, we observe an approximately linear relationship between sodium inactivation and potassium inactivation and thus set $h = 1-n$.   

The remaining gating variables are $n$, $a$, and $b$.  Their dynamics are described by equations of the form 

\begin{align} \label{gating}
X' = \phi_X \frac{X_\infty(V) - X}{\tau_X (V)}, ~~~~~ X = n, a, b.
\end{align}
The
voltage-dependent steady-state functions are of the 
form $X_\infty(V) = 1 / (1+e^{(X-\theta_X)/\sigma_X} )$. 
 For the potassium activation variable, $n$, we assume that 
$\phi_n = 0.75,\, \theta_n=-32$ and $\sigma_n=-8$. 
The time-scale for the $n$ variable is voltage-dependent: $\tau_n(V) = 1+100/(1+e^{(V+80)/26})$.
Similar to the model presented in \cite{Rush1995}, we assume that
$ \phi_a=1,\, \theta_a=-50$ and $\sigma_a=20$  for A-type potassium
activation, and $\phi_b=1,\, \theta_b=-70$ and $\sigma_b=-6$ for A-type potassium inactivation.
The time-scales for the A-type current are constants: $\tau_a = 2~\mbox{ms}$ and $\tau_b = 150~\mbox{ms}$. 

Inputs to the model include synaptic excitation ($I_{Syn,E}$) and inhibition ($I_{Syn,I}$).
Excitatory current is $I_{Syn,E} = g_{Syn,E} s_{E} (V-V_{E})$ and
inhibitory current is given by an analogous equation.  
The maximal excitatory and inhibitory conductances ($g_{Syn,E}$ and
$g_{Syn,I}$) are parameters that we vary in simulations. 
 The reversal potentials are $V_{E}=0~\mbox{mV}$ for excitation and
 $V_{I}=-85~\mbox{mV}$ for inhibition.  
The gating variables, $s_E$ and $s_I$, reset to one at the time of a synaptic
event and decay with an exponential time-course. 
That is, the excitatory gating variable is defined as

\begin{align}
s_{E}(t) =   \begin{cases}  1 \quad &\mbox{ if } t = t_E  \\  e^{-\beta_E(t-t_E)}  \quad &\mbox{ if } t > t_E \end{cases}
\label{eq:syn}
\end{align}
where $t_{E}$ is the time of the most recent excitatory event and the decay time constant is $\beta_E=0.2~\mbox{ms}^{-1}$.  
A similar equation holds for the inhibitory gating variable $s_I$, but with a time constant $\beta_I = 0.18~\mbox{ms}^{-1}$. 
Excitatory event times are randomly distributed according to a homogeneous Poisson process with rate $r_E$.  
Inhibitory event times are periodic with rate $r_I$.  
Our choice of these input patterns simplifies some of our mathematical analysis.  In addition, our choice of periodic inhibitory events was motivated by the design of {\it in vitro} experiments, 
presented in~\cite{Chen2016}, in which inhibitory interneurons were activated periodically using optogenetic techniques.
We vary the values of the rate parameters ($r_E, r_I$) in our investigations.

\subsection*{Multi-compartment neuron model}

In some simulations we augment the one-compartment (point neuron)
model by attaching additional
compartments that represent a dendritic process. We assume that
the dendrite consists of nine equally-sized compartments. 
Moreover, the neuron receives inhibitory input at its soma (the first compartment) and
excitatory input at a dendritic compartment.

Voltage in the first compartment (soma) is denoted $V_1$ and is given by Eq~\ref{eq:OneCptODE} with synaptic excitation removed and with a new term representing the flow of current between compartments (axial current):

\begin{align}
 C V_1' &= - I_{L,1} - I_K - I_A - I_{Na} -  I_{Ax,1} - I_{Syn,I}. 
\end{align}
The remaining dendritic compartments do not include potassium, A-type potassium, or sodium currents, and thus $V_j$ for $2\leq j \leq 10$ follows the linear dynamics of a passive cable:

\begin{align}
 C V_j' = \begin{cases}& - I_{L,j} -  I_{Ax,j} - I_{Syn,E}  \quad \mbox{at location of
          excitatory inputs}  \\
&- I_{L,j} -  I_{Ax,j}   \qquad \qquad ~~ \mbox{ at other locations}.
\end{cases}
\end{align}
Leak conductance in the dendrite compartments is $g_L = 0.1~\mbox{mS/cm}^2$ (one-tenth the value in the first compartment).  Axial current is
\begin{align}
I_{Ax,j} = \begin{cases} 
g_{Ax} (V_1 - V_2) &\qquad \mbox{ for } j=1 \\
g_{Ax} (-V_{j-1} + 2 V_j - V_{j+1}) &\qquad \mbox{ for } 2 \leq j \leq 9 \\
g_{Ax} (V_{10} - V_9) &\qquad \mbox{ for } j=10
\end{cases}
\end{align}
where $g_{Ax} = 10~\mbox{mS/cm}^2$.

Input currents are defined in a manner identical to inputs in the one-compartment model.  Excitatory and inhibitory gating variables follow Eq~\ref{eq:syn}. Excitatory synaptic event times are drawn from a homogeneous Poisson process with rate $r_{E}$ and inhibitory synaptic event times are periodic with rate $r_I$.  These constants, as well as synaptic input strengths ($g_{Syn,E}$, $g_{Syn,I}$) and the compartment targeted by the excitatory inputs, are parameters we vary in our investigations.

\subsection*{Computations}

We simulated the point-neuron and multi-compartment neuron models using software written in the C computer programming language.  We integrated differential equations using the fourth order implicit Runge-Kutta method available in the GNU scientific library.   We also simulated the one-compartment model and a reduced model version of the one-compartment using XPPAUT, and performed bifurcation analysis of these models using the AUTO feature of XPPAUT~\cite{Ermentrout2002}.

\section*{Results}

\subsection*{Examples of divisive and subtractive inhibition}
\begin{figure}[!h]
\centering
\includegraphics[width=\textwidth]{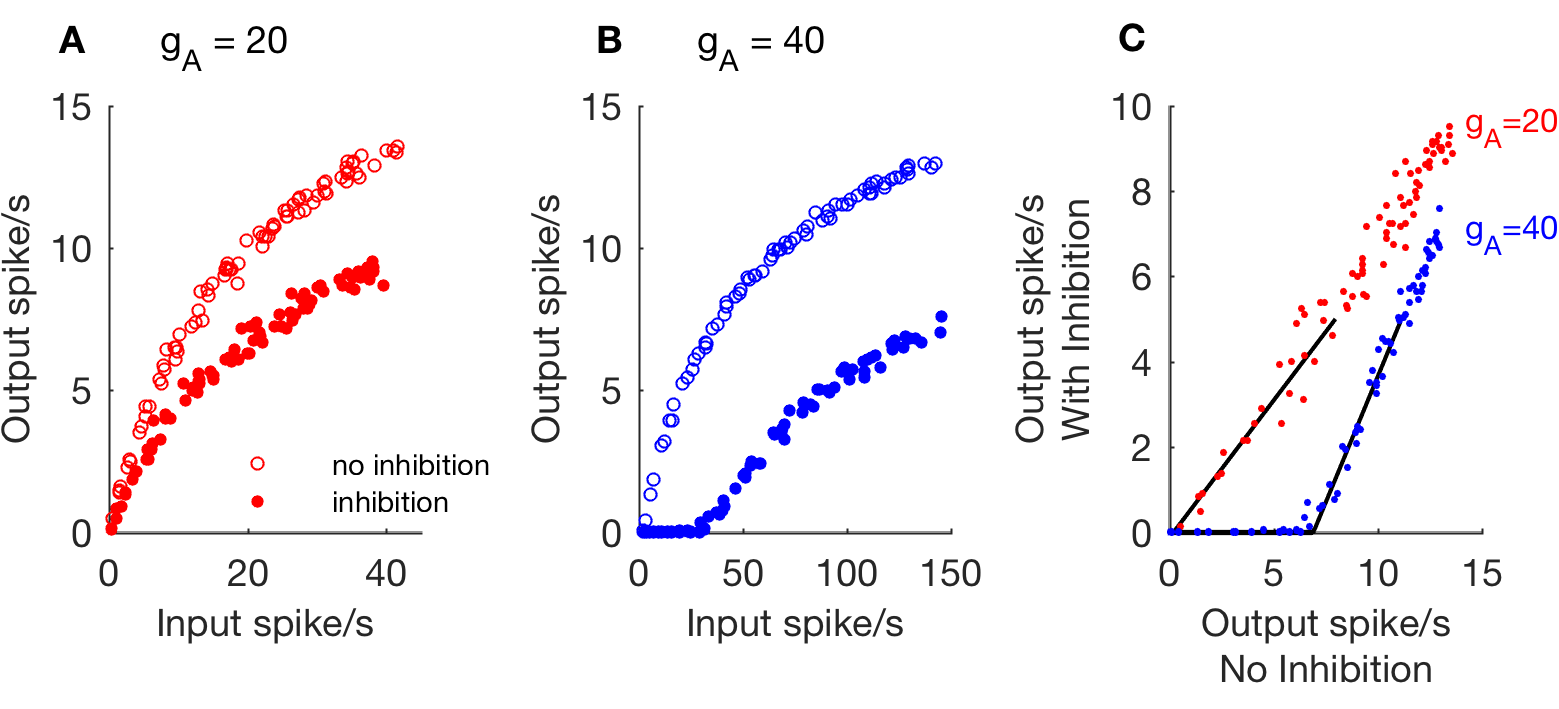}
\caption{Examples of divisive and subtractive effects of inhibition in the one-compartment model.
{\bf A, B:} Output firing rates as a function of excitatory input rate, computed from simulations without inhibition (empty circles, $g_{Syn,I}=0$) and with inhibition (filled circles, $g_{Syn,I}=1 \mbox{ and } r_I=50\mbox{ Hz}$).  Excitatory synaptic strength is $g_{Syn,E}=0.5$.
In {\bf A:} Divisive rescaling of the input/output relation with $g_A=20$.
In {\bf B:} Subtractive shifting of the input/output relation with $g_A=40$.
{\bf C: } Data from {\bf A} and {\bf B} are replotted with output firing rates in the absence of inhibition on the ordinate and output firing rates in the presence of inhibition on the abscissa.  Threshold-linear functions are fit to simulation data (black lines).  Rightward shift of threshold-linear function for $g_A=40$ is characteristic of subtractive inhibition.
}
 \label{fig:DivSub1}
\end{figure}

We first study the relationship between excitatory input rate ($r_E$)
and firing output rate ($r_{out})$ of the one-compartment model.
In Fig~\ref{fig:DivSub1}A, we plot examples of this input/output
relationship for simulations without inhibition (empty circles,
$g_{Syn,I}=0$) and with inhibition (filled circles, $g_{Syn,I}=1$).  The A-channel
conductance in these simulations is $g_A = 20~\mbox{mS/cm}^2$.
For these parameters, we observe that inhibition reduces the model neuron's output firing rate, but the neuron continues to fire in response to arbitrarily low input rates.

An additional way to view the effect of inhibition is to plot output firing rates in the presence of inhibition as a function of output firing rates in the absence of inhibition, as we have done in Fig~\ref{fig:DivSub1}C.  There is a roughly linear relationship between these output firing rates, which we describe by fitting these data with a threshold-linear function of the form 
\begin{align}
y = [ m(x-x_0) ]_+
\end{align}
where the symbol $[\cdot]_+$ indicates we set $y=0$ if the argument $m(x-x_0)$ is negative.  We obtain the slope parameter $m$ and the $x$-intercept parameter $x_0$ by applying a curve-fitting procedure (using the \emph{fminsearch} command in MATLAB) to the portion of data for which the output firing rate in the presence of inhibition is less than five spikes per second.
In this example, inhibition affects the value of the slope parameter $m$, but the value of $x_0$ is nearly zero.  
We identify responses with these characteristics as cases in which the effect of inhibition is \emph{divisive}.

In Fig~\ref{fig:DivSub1}B, we increase the A-channel conductance to
$g_A=40~\mbox{mS/cm}^2$.  We observe that inhibition has a
different effect on the input/output curve in these simulations.  In the presence of
inhibition (filled circles), there is now a non-zero value of the
input rate below which the 
neuron model does not spike ($r_{out} = 0$ for $r_E  \lessapprox 30$).  
Moreover, when we view the relationship between output firing rates with and without inhibition in Fig~\ref{fig:DivSub1}C, we observe a rightward shift of the threshold-linear function fit to these data (positive-valued $x$-intercept).  We identify responses with these characteristics as cases in which the effect of inhibition is \emph{subtractive}.

\subsection*{Parameter study: Inhibition is subtractive for strong A-current conductance or weak excitatory conductance}

\begin{figure}[!h]
\centering
\includegraphics[width=\textwidth]{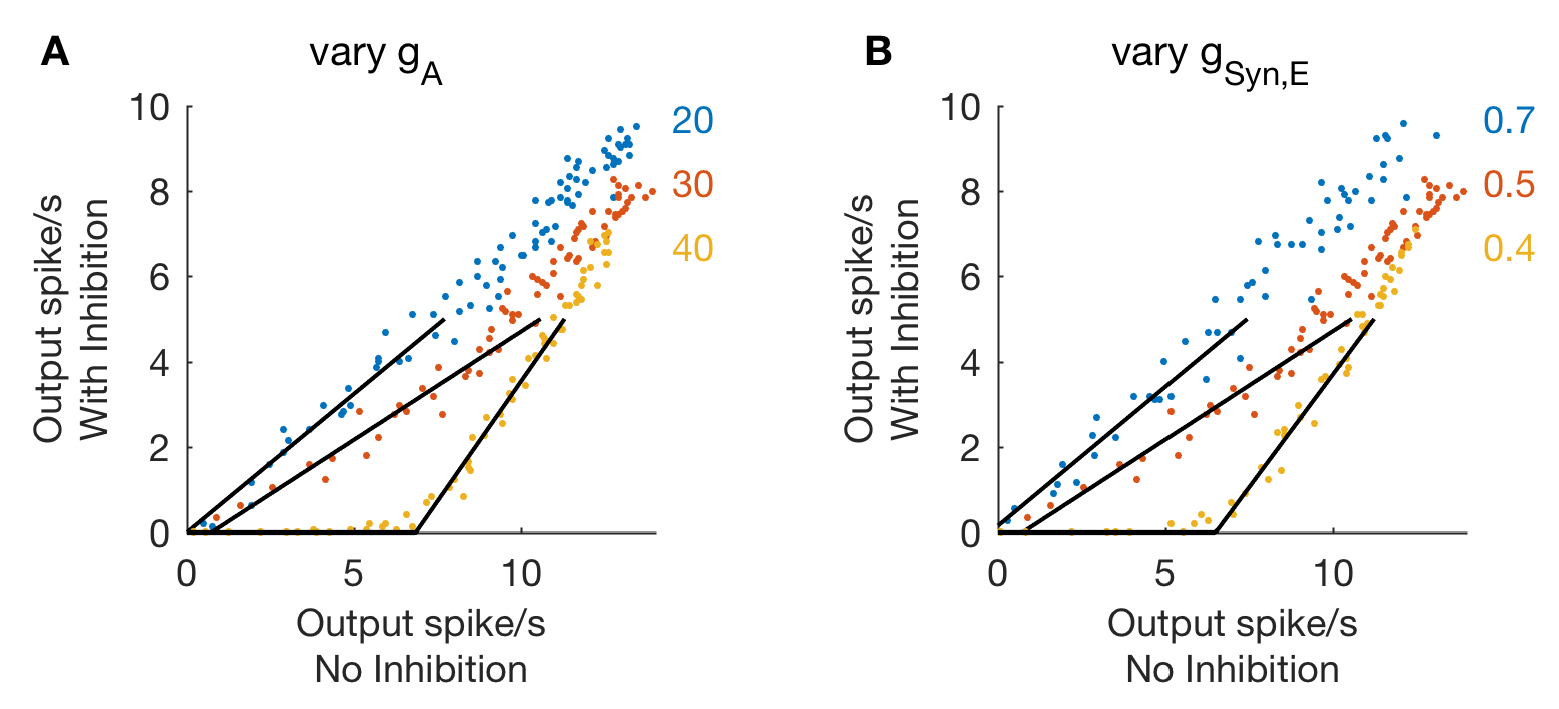}
\caption{Inhibition is subtractive for large A-channel conductance or weak synaptic excitation.
{\bf A, B:} Firing rates computed from simulations with inhibition ($g_{Syn,I}=1, r_I=50\mbox{ Hz}$, abscissa) plotted as a function of firing rates computed from simulations without inhibition ($g_{Syn,I}=0$, ordinate).  
In {\bf A:}  Three values of A-channel conductance are compared ($g_A=20, 30, 40$) with synaptic excitation strength fixed at $g_{Syn,E}=0.5$. 
Inhibition is subtractive for large $g_A$ evident in the rightward shift of the threshold-linear relationship between firing rates for $g_A=40$.
In {\bf B:}  Three values of synaptic excitation strength are compared ($g_{Syn,E}=0.4, 0.5, 0.7$) with A-channel conductance fixed at $g_A=30$. 
Inhibition is subtractive for weaker excitation, evident in the rightward shift of the threshold-linear relationship between firing rates for $g_{Syn,E}=0.4$.
}
\label{fig:threshLinear1}
\end{figure}

We identify two parameters in the one-compartment model that are key factors in determining whether inhibition has a divisive or subtractive effect on firing rate responses: the A-channel conductance ($g_A$) and the excitatory synaptic conductance ($g_{Syn,E}$).   In Fig~\ref{fig:threshLinear1}A we show a set of threshold-linear functions computed using $g_A=20, 30\mbox{ and } 40$, and synaptic excitation strength fixed at $g_{Syn,E}=0.5$.  The transition from divisive to subtractive inhibition is evident in the rightward shift of these threshold-linear functions with increasing values of $g_A$.  This transition occurs, for this parameter set, for $g_A\approx33$, a point we investigate in more detail below, with simulations and phase plane analysis.

In Fig~\ref{fig:threshLinear1}B, we show a set of threshold-linear functions with $g_A=30$ fixed, but now varying the value of $g_{Syn,E}$ from 0.4 to 0.7.  
The stronger excitatory inputs ($g_{Syn,E} = 0.5, 0.7$) cause inhibition to have a divisive effect, while the weaker excitatory input ($g_{Syn,E}=0.4$) causes inhibition to have a subtractive effect.  
In these simulations, we do not vary the parameters associated with inhibition.  They are $g_{Syn,I}= 1$ and $r_I=50~\mbox{Hz}$.  

From these simulations, we conclude that the effect of inhibition on
firing rates in the one-compartment model can switch from divisive to
subtractive for sufficiently strong A-current conductance or
sufficiently weak excitatory synaptic conductance.  In the parameter
plane of $g_A$ and $g_{Syn,E}$, then, there is a boundary that
separates parameter sets that produce divisive inhibition from
parameter sets that produce subtractive inhibition.  We map this
boundary by performing simulations 
throughout the $(g_A,\, g_{Syn,E})$ parameter space.  For each simulation, we fit threshold-linear functions to characterize the relationship between output firing rates in the presence and absence of inhibition.  We then find the smallest value of $g_A$ for which the $x$-intercept of the threshold-linear function is right-shifted by more than two spikes per second and label this as boundary between subtractive and divisive inhibition.

\begin{figure}[!h]
\centering
\includegraphics[width=\textwidth]{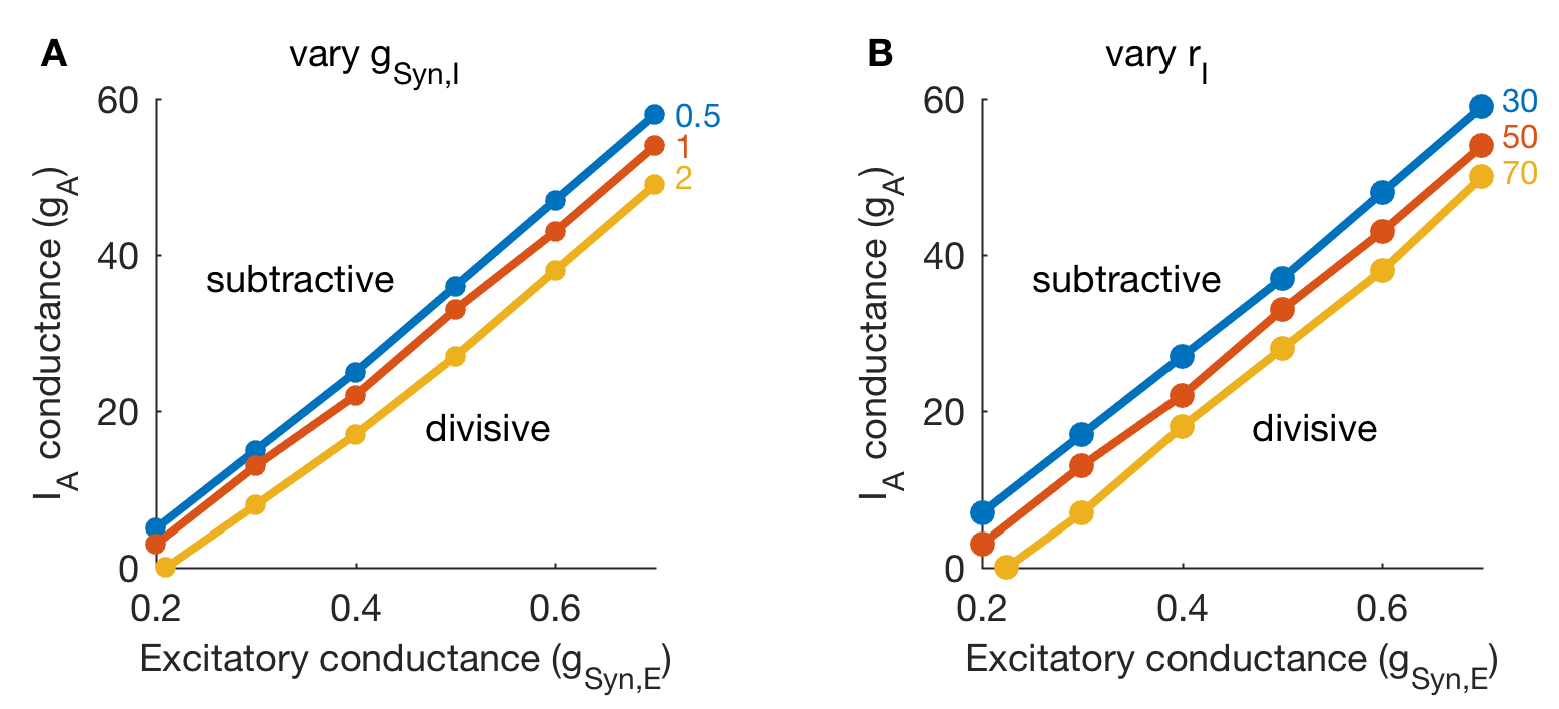}
\caption{Boundary between subtractive and divisive inhibition in $(g_{Syn,E},g_A)$ parameter space.
{\bf A, B:} For each parameter set, we fit threshold-linear functions to characterize the relationship between output firing rates in the presence and absence of inhibition.  Dots in each panel identify the smallest value of $g_A$ (for a given parameter set) at which inhibition is subtractive. %($x$-intercept of threshold-linear function right-shifted by two spikes per second).  
In {\bf A:} We vary inhibition strength ($g_{Syn,I}=0.5, 1, 2$) and keep inhibition rate fixed at 50~Hz. 
In {\bf B:} We vary inhibition rate ($r_I = 30, 50, 70$~Hz) and keep inhibition strength fixed at $g_{Syn,I}=1$.
The values of $g_A$ that define the boundary between subtractive and divisive inhibition decrease with increases in either inhibition parameter ($g_{Syn,I}$ or $r_I$). }
\label{fig:DivSubBoundary1}
\end{figure}

In Fig~\ref{fig:DivSubBoundary1}, we show the results of this parameter exploration.  We performed these simulations and classification procedure for several values of inhibition conductance strength (varying values of $g_{Syn,I}$, in Fig~\ref{fig:DivSubBoundary1}A), and for several values of inhibition rate (varying values of $r_I$, in Fig~\ref{fig:DivSubBoundary1}B).  The lines in each panel separate parameter regions for which inhibition is divisive (lower right corners in each panel) from parameter regions in which inhibition is subtractive.  This confirms our earlier observation that the effect of inhibition is subtractive if A-channel conductance is sufficiently strong or excitatory inputs are sufficiently weak.

These simulations also demonstrate that inhibition parameters
modify (weakly) the location of the boundary between divisive and
subtractive 
inhibition in the $(g_A,\,g_{Syn,E})$ parameter plane.  Stronger inhibition
(either through larger $g_{Syn,I}$ or larger $r_I$ values) decreases the
 portion of the $(g_A,\,g_{Syn,E})$ parameter plane in which inhibition has a divisive effect on firing rate responses.

\subsection*{Analysis of a reduced one-compartment model}

We use mathematical analysis to derive the parameter regions in which the model exhibits either a divisive or subtractive 
response to inhibition.  We begin by considering a reduced model in which activation of the A-current is instantaneous; that is, $a = a_\infty(V)$. Later, we discuss
how the model's response to inhibition may change if this assumption
does not hold.

\subsubsection*{Excitability analysis using fast/slow dissection}
\begin{figure}[!h]
\centering
\includegraphics[width=\textwidth]{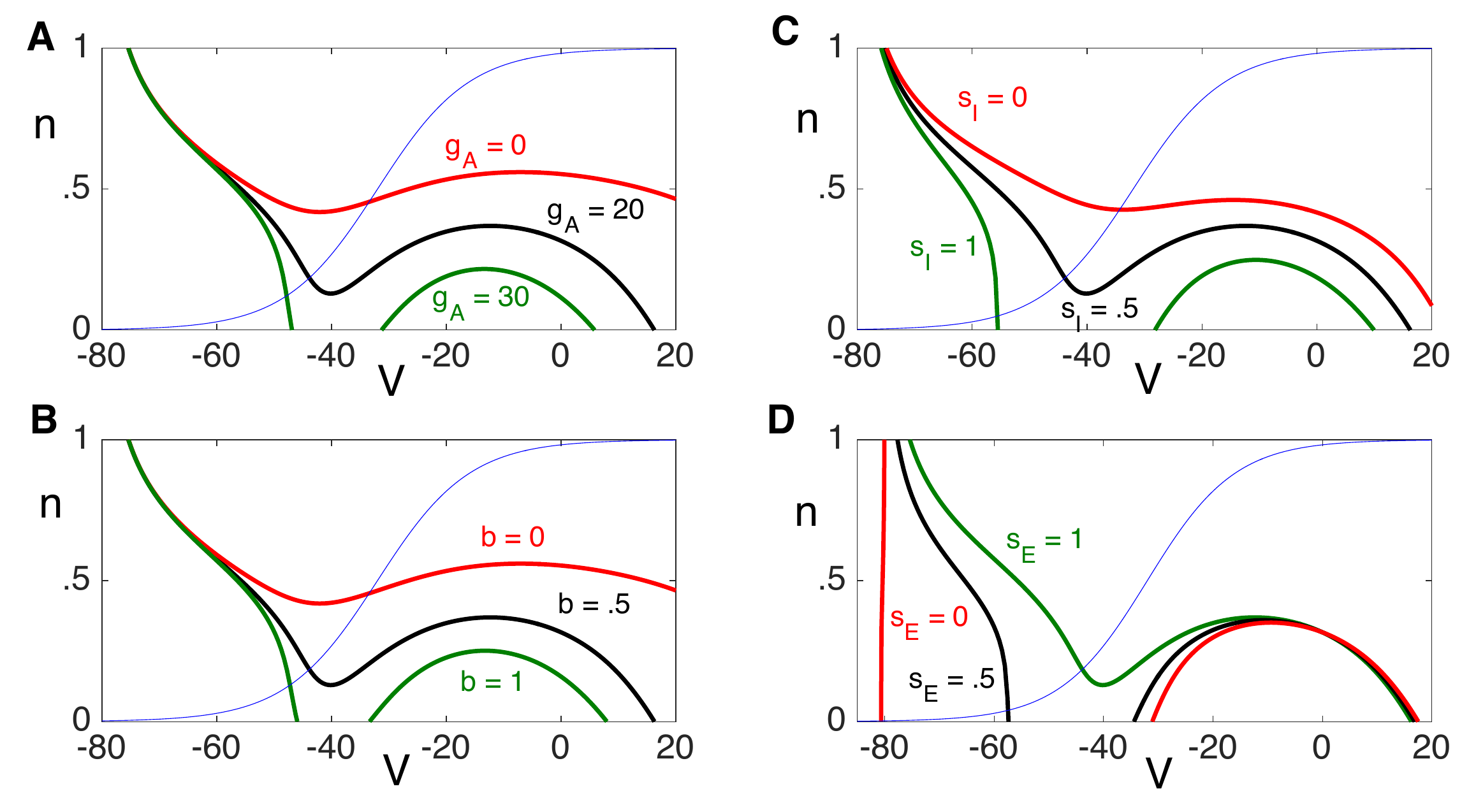}
\caption{Dependence of the $V$-nullcline on {\bf A:}~$g_A$, {\bf B:}~$b$,
  {\bf C:}~$s_I$ and {\bf D:}~$s_E$. Default values of the parameters are $g_A = 20, b=.5, s_I=.5$
  and $s_E=1$.  Moreover, $g_{Syn,E} = 3$ and $g_{Syn,I} = 5.$ Thin blue line is $n_\infty(V)$, the $n$-nullcline.}
\label{knees}
\end{figure}

A first step in the analysis is to determine under what conditions the
neuron will fire an action potential in response to an excitatory
input. An important distinction between divisive and subtractive
inhibition is whether the neuron can respond to arbitrarily low
excitatory  input rates. 
We  analyze the model by viewing solutions in the $(V,n)$ phase plane.
A major challenge in this approach is that the phase plane depends 
not only on the other dependent variable, $b$, but on the values of
the synaptic inputs, $s_E$ and $s_I$. Suppose, for the time being,
that these variables
 are fixed constants. Fig~\ref{knees} shows  phase planes for different
 values of these constants. In particular, it illustrates how the
 $V$-nullclines change as $b,\,s_E,\,s_I$ 
and the parameter $g_A$ changes. In each case, the $V$-nullcline has
left, middle and right branches,
while the $n$-nullcline is a monotone increasing
function. Note that values of $n$ along the
 $V$-nullcline are  increasing functions of $s_E$ and  decreasing
 functions of $b, s_I$ and $g_A$.

We consider $V$ to be  a fast variable and $n$ and $b$ to be slow
variables. During silent or active phases,  solutions lie on,
respectively, the left 
or right branch of the $V$-nullcline corresponding to
values of $b,\,s_E$ and $s_I$. The jumps up and down of action
potentials 
correspond to horizontal transitions between  left and right branches in the phase plane.

Now suppose that the solution initially lies in the silent phase along the left branch of some $V$-nullcline
and there is an excitatory synaptic input. Then $s_E$ will immediately
jump from $s_E = 0$ to $s_E = 1$, resulting in an immediate change in
the $V$-nullcline, 
as shown in Fig~\ref{knees}D.  If $(V(0),n(0))$ lies below the left knee of this new $s_E = 1, ~V$-nullcline, then 
the solution will jump to the right branch of the new $V$-nullcline,
resulting in an action potential. However, if $(V(0),n(0))$ lies above
this left knee, then the 
solution will jump to the left branch of the new $V$-nullcline; that is, the solution will not respond to the excitatory input with an action potential.

This discussion helps to explain when the neuron will or will not fire
an action potential in response to an excitatory input. In particular,
there are two reasons 
why the neuron may not respond. The first reason is that the neuron
may be in a refractory period. That is, if the excitatory input
arrives shortly after the neuron 
has previously fired, then the K$^+$ activation variable, $n$, may not
have had enough time to recover and evolve below the left knee of the
$s_E = 1$ cubic
 nullcline. The second reason is that the inhibitory
 input, $s_I$, may be too strong and the left knee of the $s_E=1$
 cubic may lie below the $n=0$ 
axis. In this case, the neuron will not respond even though there has been a long time since the preceeding excitatory input.

\paragraph{Estimate of $b$:}
One difficulty is that the  cubic nullclines depend on the time-dependent variable $b$. 
For the analysis, we  replace $b$ by its average value along solutions,  which we estimate as follows. 
A key observation  is that the average value of  $b$ can be well approximated by
considering a model consisting only of synaptic inputs and the leak current. In 
particular, to compute the average value of $b$  we may ignore the voltage dependent
Na$^+$,  K$^+$ and A-currents. 

\begin{figure}[!h]
\centering
\includegraphics[width=\textwidth]{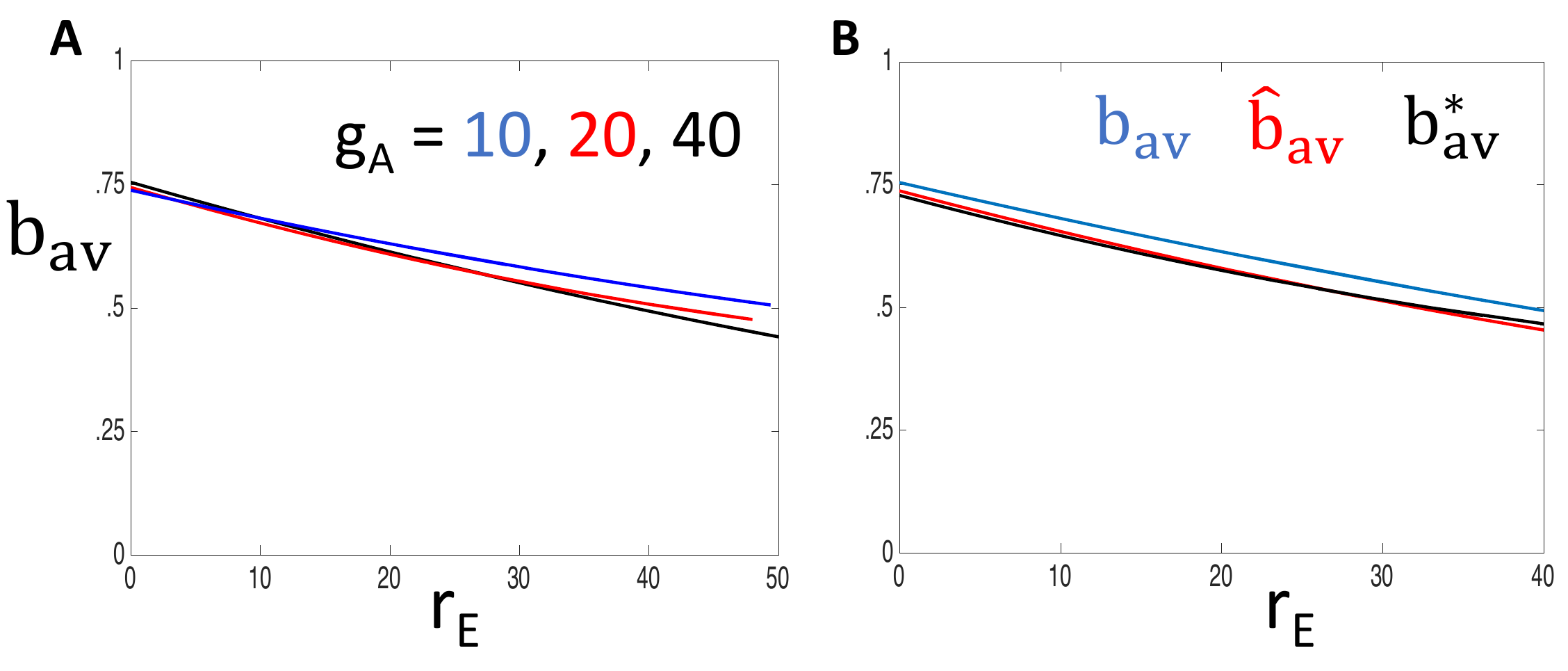}
\caption{Approximation of $b$ (a slow variable) by its average value.  {\bf A:} Plots of $b_{av}$ vs. $r_E$ for different values of
  $g_A$.
{\bf B:} Plots of $\hat{b}_{av}(r_E)$, $b^*_{av}(r_E)$ and
$b_{av}(r_E,g_A)$ with $g_A=40$. In both panels, $g_{Syn,E}=3,
g_{Syn,I}=5$ and $r_I =$ 50 Hz. }
 \label{bav}
\end{figure}

To justify this claim, we first simulate the full model for different values of $g_A$ and  $r_E$,
and then compute the average value of $b$ along solutions, which we denote as $b_{av}(r_E,g_A)$.
Fig~\ref{bav}A shows plots of $b_{av}$ versus $r_E$ for different values of $g_A$.
Note that $b_{av}$ depends only weakly on $g_A$ and is a decreasing function of $r_E$.

We next repeat  these simulations with $g_A = g_K = g_{Na} =
0$. Actually, we perform the simulations twice: once with purely
excitatory inputs ($r_I = 0$) and 
then with purely inhibitory inputs ($r_E = 0$). This gives rise to a curve $\hat{b}_E(r_E)$ and a constant $\hat{b}_I$.
We then let
$$\hat{b}_{av}(r_E) =  \hat{b}_I + \hat{b}_E(r_E) - \hat{b}_E(0).$$
Fig~\ref{bav}B shows that $\hat{b}_{av}(r_E)$ gives an excellent approximation of $b_{av}(r_E,g_A)$.

We can simplify the anaysis further by considering a fast/slow
reduction with $V$ as the fast variable. If $g_A = g_K = g_{Na} = 0$, 
then $V$ satisfies the {\it linear} equation:
$$ V' = -g_L(V-V_L) - g_{Syn,E}\,s_E\,(V-V_{Syn,E}) - g_{Syn,I}\,s_I\,(V-V_{Syn,I}).$$
Treating $V$ as a fast variable, we set the right hand side of this equation equal to $0$ and solve for $V$ to obtain
\begin{align}
V = \frac{g_L\,V_L + g_{Syn,E}\,s_E\,V_{Syn,E} + g_{Syn,I}\,s_I\,V_{Syn,I}} {g_L + g_{Syn,E}\,s_E + g_{Syn,I}\,s_I}.
\end{align}
As before, we treat excitatory inputs and inhibitory inputs separately. Let
$$
V_E = \frac{g_L\,V_L + g_{Syn,E}\,s_E\,V_{Syn,E}}{g_L + g_{Syn,E}\,s_E } ~~ \mbox{and} ~~~
V_I = \frac{g_L\, V_L  + g_{Syn,I}\,s_I\,V_{syn,I}} { g_L  + g_{Syn,I}\,s_I}.
$$
Then let $b^*_E(r_E,t)$ and $b^*_I(t)$ be solutions to
 Eq~\ref{gating} with $X=b$, and $V = V_E$ and $V=V_I$, respectively.
Finally, let $b^*_E(r_E)$ and $b^*_I$ be the average values of $b^*_E$ and $b^*_I$ along solutions and set 
$$ b^*_{av}(r_E) = b^*_I +  b^*_E(r_E) - b^*_E(0). $$
Fig~\ref{bav}B shows that $b^*_{av}(r_E)$ gives an excellent approximation of $\hat{b}_{av}(r_E)$. In the analysis that follows, we replace the dependent variable $b$
by  constants $b^*_{av}(r_E)$. 

\paragraph{The left knee:}
Our previous discussion emphasized the importance of the left knees of the cubic-shaped $V$-nullclines when $s_E = 1$.
 If this left knee lies below the $n=0$ axis, then the neuron cannot spike in response to an excitatory input. 
For the reduced model, with $a=a_{\infty}(V)$, this left knee depends on the values of $s_I, b$ and $g_A$. 
Here, we replace $b$ by $b^*_{av}(r_E)$ and  denote the value of $n$
at the left knee as $N_{lk}(s_I, r_E, g_A)$. 
We compute the positions of the left knees using XPPAUT.

\paragraph{Identification of inhibition as divisive or subtractive:}
By computing the position of the left knee, we can determine for which parameter values inhibition will have a divisive effect and for which values it will have a subtractive effect.
Let $P_I$ be the period of inhibitory inputs. Since $s_I' = -\beta_I s_I$ between inhibitory inputs, it follows that
\begin{align}
s_I(t) ~ \geq ~ e^{-\beta_I P_I} ~\equiv~ \sigma_*
\end{align}
for all $t$. 
 
\begin{figure}[!h]
\centering
\includegraphics[width=\textwidth]{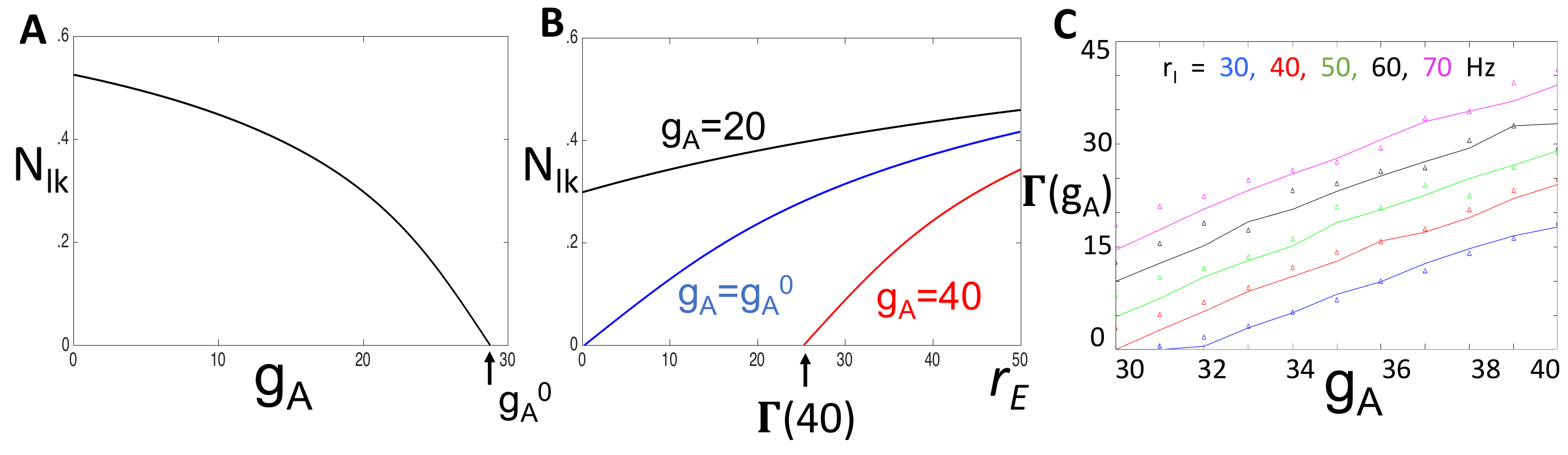}
\caption{Identifying $g_A$ value at the point of transition between divisive and subtractive inhibition.  Dependence of the left knee, $N_{lk}(s_I, r_E, g_A)$, on {\bf
    A:} $g_A$ with $s_I = \sigma_*$ and $r_E = 0$; and {\bf B:} $r_E$
  with $s_I = \sigma_*$ and different values of $g_A$. Here, $g_A^0 =
  28.85$.
{\bf C:} Theoretical calculation of $\Gamma(g_A)$ (solid lines)  and 
measured values of the minimum input rate at which solutions of the
reduced model exhibits non-zero output firing rates (small triangles)
for different values  of the inhibitory input rate, $r_I$.}
 \label{leftknee}
\end{figure}

Fig~\ref{leftknee}A shows plots of $N_{lk}$ versus $g_A$ when
$r_E=0$ and $s_I = \sigma_*$. Note that $N_{lk}$ is a decreasing
function of $g_A$ and there exists  $g_A^0$ such that 
$N_{lk}(\sigma_*, 0, g_A^0) = 0$. Since $N_{lk}$ is an
increasing function of $r_E$, it follows that if $g_A < g_A^0$, then 
$N_{lk}(\sigma_*, r_E, g_A) > 0$ for all $r_E > 0.$
Since $N_{lk}$ is a continuous function of $s_I$, it follows that if
$g_A< g_A^0$ and  $s_I > \sigma_*$ with $s_I - \sigma_*$ sufficiently small,
then 
$N_{lk}(s_I, r_E,  g_A) > 0$ for all $r_E$.
In this case, the neuron is able to respond to arbitrarily low firing rates and we expect the response
to inhibition to be divisive. 

Fig~\ref{leftknee}B shows plots of $N_{lk}$ versus $r_E$ for
different values of $g_A$ when $s_I = \sigma_*$
Note that if $g_A > g_A^0$, then there exists a critical value of $r_E$, which we denote by
$\Gamma(g_A)$, such that if  $r_E < \Gamma(g_A)$, then
$N_{lk}(\sigma_*,r_E,g_A) < 0$;
 if $r_E > \Gamma(g_A)$, then $N_{lk}(\sigma_*,r_E,g_A) > 0$.
Recall that $N_{lk}(s_I,  r_E, g_A)$ is a decreasing function of $s_I$.
Hence, if $g_A > g_A^0$ and $r_E < \Gamma(g_A)$, then $N_{rk}(s_I, r_E, g_A) < 0$ for
all $s_I > \sigma_*$. Moreover, if $g_A > g_A^0$ and $r_E >
\Gamma(g_A)$ then
$N_{rk}(s_I,r_E,g_A) > 0$ for  $s_I > \sigma_*$ with $s_I - \sigma_*$ sufficiently small.
Hence, if $g_A > g_A^0$, the neuron is not able to respond to arbitrarily low firing rates and we expect the response
to inhibition to be subtractive. Given $g_A > g_A^0$, the minimum
input rate at  which the neuron can respond is $\Gamma(g_A)$. Fig~\ref{leftknee}C compares our theoretical calculation of $\Gamma(g_A)$ to measured values of the minimum input rate at which the reduced model exhibits non-zero output firing rates (theoretical approximation shown with solid line, simulation results shown with symbols).  We report this comparison between theory and computation for values of inhibitory input rate ($r_I$) that vary from 30~Hz to 70~Hz in increments of 10~Hz.

\subsubsection*{Output rate approximation using dead-time modified Poisson process}
The previous analysis was concerned with small output firing rates.
Here we consider output rates bounded away from zero. For this
analysis, we first consider the model without inhibition and discuss
the impact of the refractory period on the neuron's output rate. We
then discuss the impact of inhibitory inputs. 

\paragraph{Formulation of output rate as dead-time modified Poisson process:}
In our simulations, we typically use excitatory inputs that are sufficiently strong so that, in the absence of refractory effects and inhibition,
a single excitatory input event triggers an output spike.  
Thus, at low input rates and with $g_{Syn,I}=0$, we expect generation of output spikes to replicate the sequence of input excitatory events. 
Specifically, the output spike train will follow a homogeneous Poisson process with the same rate as excitatory events: $r_{out} = r_E$.

For high input rates, a refractory effect prevents the neuron from firing on a one-to-one basis with each input event.  
A simple approximation of the input/output relationship in this ``high input rate'' regime is to assume there is a fixed period of duration $R~\mbox{ms}$ during which the neuron cannot fire, and therefore we say
\begin{align}
\mbox{Input event at time t} \rightarrow \begin{cases} \mbox{Output spike if previous spike occurred} > R \mbox{ in the past}  \\ \mbox{No output spike otherwise.} \end{cases}
\end{align}
Since input events are drawn from a homogeneous Poisson process, the output events under this approximation follow a
dead time modified Poisson process with dead time (i.e. refractory period) $R$.  The input/output firing rate relationship is then given by
\begin{align}
r_{out} = \frac{r_E}{1 + r_E R} 
\label{eq:PoissonE}
\end{align}
where rate is defined as the expected number of events a time interval, divided by the duration of that interval~\cite{Muller1974}.

\paragraph{Definition of the firing threshold:}
The approximation of $r_{out}$ introduced in Eq~\ref{eq:PoissonE} does not incorporate any effect of inhibition.
To incorporate the effects of inhibition into the formulation,
we presume that excitatory input events will produce output
spikes (following the dead-time modified Poisson process, as described above) unless the
excitatory event occurs at a time when synaptic inhibition is
sufficiently strong to prevent spiking.  
In other words, we define a \emph{firing threshold} $\theta=\theta(r_E,g_A)$ to
be the value of the inhibitory conductance $s_I$ for which an 
excitatory input will evoke a spike only if $s_I<\theta$. 
For now, we assume that $\theta$ is known. 
Later, we describe how $\theta$ can be computed.

Suppose that inhibitory events occur periodically with period
$P_I$. At the onset of each inhibitory event, the 
 inhibition conductance $s_I$  increases to 1 and then exponentially
decays with time constant $1/\beta_I$. 
It follows that the fraction of time that 
that $s_I<\theta$ is given by
\begin{align}
\rho= 1 + \frac{\log \theta}{P_I \beta_I}.
\end{align}
By the definition of the firing threshold, this is the fraction of
time that the neuron responds to excitatory input. 
Hence, we modify the Poisson approximation of output firing rate to be
\begin{align}
\label{eq:dtinhib}
r_{out} = \frac{r_E~\rho}{1 + R r_E}.
\end{align}
To complete the analysis, it still remains to compute the firing
threshold, $\theta$, which depends on the  parameters
$r_E$ and $g_A$.

\paragraph{Computing the firing threshold:}
Our derivation of the firing threshold procedes as follows.
Let $n_{av}$ be the average value of $\{n_k: k = 1, 2, .... \}$ such
that for each $k$: i) there is an excitatory input at some time, $t_0$,
with $n(t_0) = n_k$; and ii) the neuron responds to this excitatory
input with a spike. We show below that if $n_{av}$ is known, then we can
compute the firing threshold, $\theta$, while if $\theta$ is known, then we can compute $n_{av}$.
This gives rise to two maps: $\theta = \Theta(n_{av})$ and $n_{av} =
N(\theta)$. If $(\theta^*, n^*)$ is a fixed point of these two maps,
then the firing threshold is given by $\theta=\theta^*$. In  what
follows, we make several simplifying assumptions. We compare
predictions of the analysis with simulations of the reduced model later.

Suppose that $n_{av}$ is given. Here we assume that if an excitatory
input arrives at time $t_0$ and the neuron responds to this input with a
spike, then $n(t_0)=n_{av}$. Recall that the neuron will spike in
response to an excitatory input at time $t_0$
if $(V(t_0),n(t_0))$ lies below the left knee of $s_E = 1$ nullcline, which depends on the other variables, $b$ and $s_I$.
As before,  let $b = b^*_{av}(r_E)$. The
firing threshold can then be defined as the value of $s_I$ so
that $N_{lk}(s_I, r_E, g_A) = n_{av}$. This defines the map $\theta = \Theta(n_{av})$.

Now suppose we are somehow given the firing threshold $\theta$ and
wish to estimate $n_{av}$. The first step is to determine the position
of the right knee of the $s_E = 1$ cubic, which we denote as $N_{rk}$. This
depends on $s_I, b$ and $g_A$; however, as  the
dependence of $N_{rk}$ on each of these variables is weak, we will
assume that $N_{rk}$ is a  constant.  We determine its value using XPPAUT.

We next consider the evolution of $n$ in the silent phase. Here, we
assume that $n_\infty(V)=0$ and $\tau_n(V) = \tau_0$, a
constant. Then, while in the silent phase, $n(t)$ satisfies
$n' = -\frac{\phi_n}{\tau_0}~ n $, so that
$n(t) = N_{rk} \, e^{-\frac{\phi_n \,  t}{\tau_0} }.$
If the firing threshold $\theta$ is known, then the output firing rate, $r_{out}$,
is given by Eq~\ref{eq:dtinhib}. The average interspike interval is, therefore,
$1000/r_{out}$ ms, which we assume is the average time the neuron spends in
the silent phase. In this case, the average value of $n$ is given by
$$n_{av}  = N_{rk} \,  e^{-\frac{1000\,\phi_n}{ \tau_0\,r_{out}}}.$$
This defines the map $ n_{av} =N(\theta)$.

We have now defined the two maps, $\theta = \Theta(n_{av})$ and
$n_{av}=N(\theta)$. By taking the composition of these maps, we obtain
the fixed point problem: $n= N \circ \,\Theta(n)$.  
This map is continuous in $n$, and $n$ takes values in the closed interval $[0,1]$, so there exists a fixed-point of this map~\cite{Crossley2010}.  
We use successive iterations of the bisection method to numerically
compute $n_{av}$ as the solution of the fixed point fixed point problem $n = N \circ \,\Theta(n)$, and then set $\theta = \Theta(n_{av})$.

\paragraph{Slope of output rates at onset of firing in the divisive case:}
By computing the firing threshold at $r_E = 0$, we can study the behavior of the model near the onset of firing.
In particular, if $g_A$ is sufficiently small or $g_{Syn,E}$ is sufficiently large, then inhibition is divisive and the
neuron is able to fire in response to arbitrarily low firing rates (recall Fig~\ref{fig:DivSub1}). 
Our goal here is to compute the slopes of the input/output
firing rate curves at $r_E = 0$ in the case of divisive inhibition. 

This is done by simply differentiating the firing rate approximation in
Eq~\ref{eq:dtinhib} with respect to $r_E$ and setting $r_E = 0$. This
  yields
\begin{align}
\label{eq:slope}
r_{out}'(0) = 1 + \frac{\log \theta_0}{P_I \beta_I}
\end{align}
where $\theta_0$ is the firing threshold when $r_E =  0$. 
This can be easily computed using XPPAUT as follows. Since we are considering
arbitrarily low firing rates, it follows that the average value of $n$ at times of output spikes is $n_{av} \approx 0$,
where $n_{av}$ was defined in the preceding section. 
The analysis in that section demonstrates that $\theta_0 = \Theta(0)$, i.e. it is the value of $s_I$ so that $N_{lk}(s_I,0,g_A) = 0$.

\begin{figure}[!h]
\centering
\includegraphics[width=4in,height=3in]{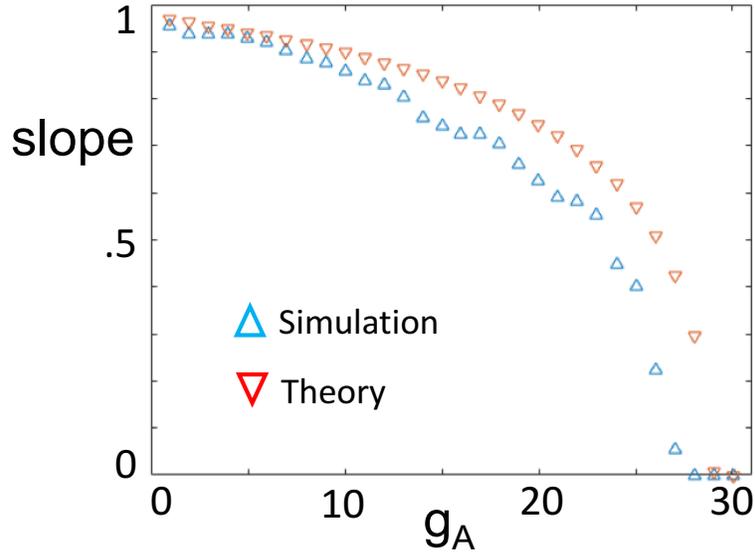}
\caption{The slope of the input/output firing
rate curves at $r_E = 0$ computed from both the theoretical prediction
Eq~\ref{eq:slope}
and simulations of the full model.}
 \label{slope}
\end{figure}

In Fig~\ref{slope}, we plot the slope of the input/output firing
rate curves at $r_E = 0$ computed from both the theoretical prediction
(Eq~\ref{eq:slope})
and simulations of the full model.
There is a tendency for the approximated value of the slope at firing onset to overestimate the slope computed in simulations of the full model.   
Nonetheless, the theory captures qualitative features of the relationship between slope at firing rate onset and A-channel conductance. 
In particular, the slope at firing rate onset decreases as $g_A$ increases, indicating gain control by inhibition is ``stronger'' with higher values of $g_A$. The value of $g_A$ at which the slope reaches zero ($g_A$ between 25 and 30) was denoted as $g_A^0$ previously; it is the critical value of $g_A$ at which the effect of inhibition switches from divisive to subtractive.

\paragraph{Approximation of output rates at arbitrary input rates:}

To more completely characterize output firing rates, we seek to extend our approximations to cases of higher output rates.
In other words, we consider the firing rate equation (Eq~\ref{eq:dtinhib}) for $r_E>0$.   
We previously described how to compute $\theta$.
With $\theta$ known, the only undetermined parameter in the firing rate equation is $R$.  We interpret this parameter as the duration of the refractory period in the model.
The value of $R$ depends on the internal dynamics of the model, strength of excitatory inputs, and other factors.  To obtain approximations for $R$, we simulated firing rate input/output curves (without inhibition) and performed a nonlinear curve-fit using Eq~\ref{eq:PoissonE} to estimate its value.
We find that $R\approx 10~\mbox{ms}$ (with some dependence 
on $g_a$) provides accurate approximations to the firing rate
functions in the absence of inhibition.

\begin{figure}[!h]
\centering
\includegraphics[width=\textwidth]{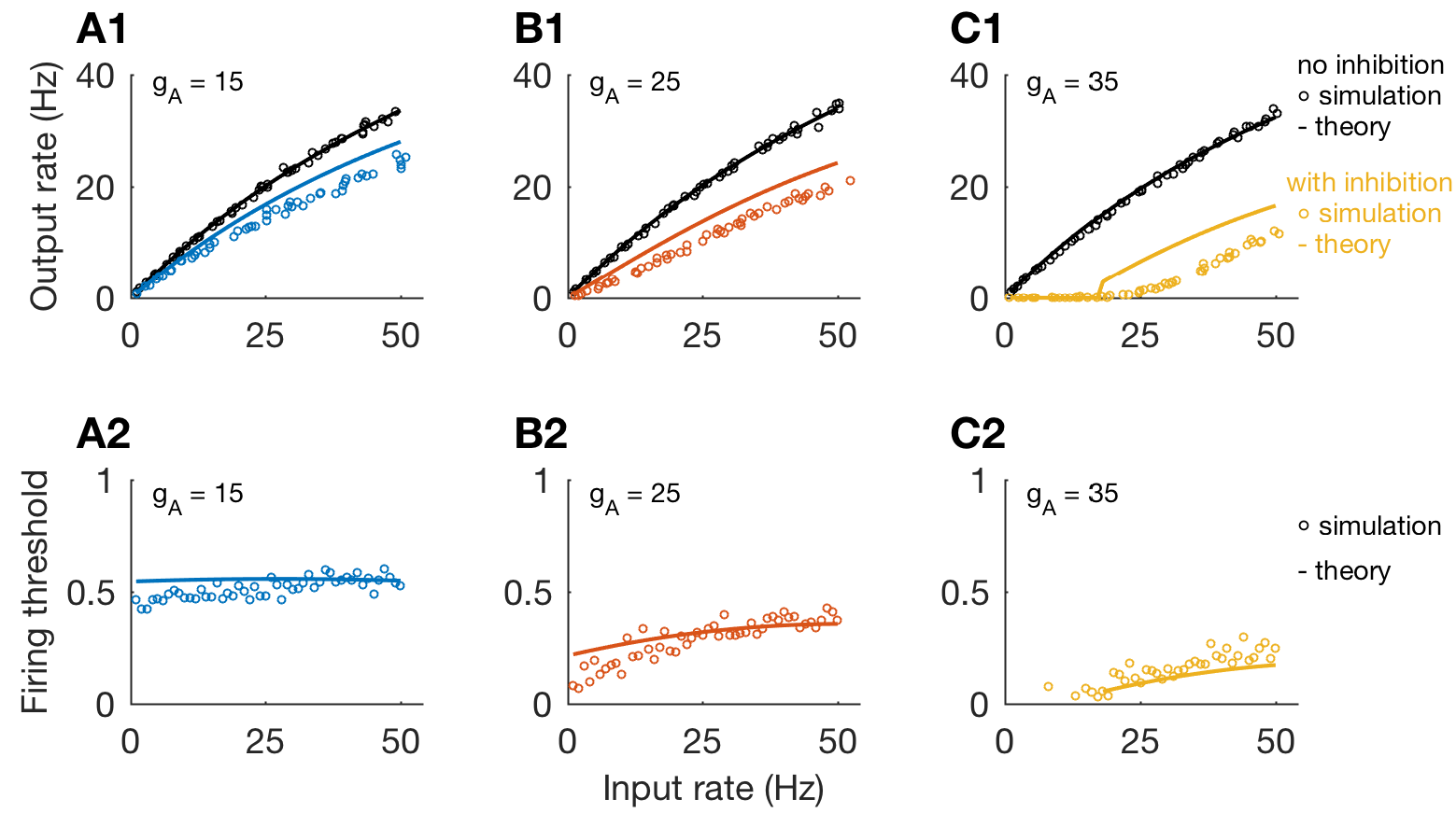}
\caption{Output firing rates approximated as dead-time modified Poisson process with firing threshold.
{\bf Top row:} Firing rate as a function of input rate obtained from simulations (circles) and theoretical approximation (lines) for three different $g_A$ values.  
Simulations and theory show transition from divisive to subtractive inhibition as $g_A$ increases.  
{\bf Bottom row:} Theoretical approximation of firing threshold $\theta$ (lines), and the largest observed values of $s_I$ for which excitatory inputs elicited spikes in simulations (circles), plotted as functions of input rate.
}
 \label{fig:firingThreshold}
\end{figure}

We compare our theoretical approximation to the firing rate of the
model neuron to simulated firing rates in
Fig~\ref{fig:firingThreshold}.  
The dead-time modified Poisson process provides a satisfactory
description of firing rate responses without inhibition (black, top
row), for various values of $g_A$ (values of $g_A$ increase from 15 to
35, in increments of 10, from left-to-right in this figure).  
When we include inhibition in these simulations, we find that the approximated firing rate (using the firing rate threshold
computation outlined above) tends to 
overestimate simulated firing rates (colors).  Nevertheless, approximations do capture
the qualitative differences in firing rate curves for divisive
inhibition (smaller values of $g_A$) and subtractive inhibition (larger values of $g_A$).

The firing threshold calculation defines $\theta$, the theoretical value of $s_I$
for which the neuron cannot produce a spike if an incoming excitatory
event arrives at a time when $s_I \ge \theta$.  We plot values of $\theta$
as colored lines in the lower row of
Fig~\ref{fig:firingThreshold}.  
To compare $\theta$ to simulations, we recorded the values of $s_I$ at
the time of every excitatory event that triggered a spike.  
Then, for each input rate (x-axis), we found the maximum value of $s_I$ for these spike-triggering excitatory events.
These maximum $s_I$ values qualitatively align with our approximations of $\theta$, further supporting the heuristic concept of a firing threshold.

\subsection*{Results for non-instantaneous A-current activation}

A key assumption in the formulation and analysis of the reduced model
is that the A-current activates sufficiently fast so that the
dynamical variable $a$ can be set to its voltage-dependent steady state value; that is, we
set $a = a_\infty(V)$.  
One effect of this change from $a$ evolving dynamically with $\tau_a= 2~\mbox{ms}$ (the full model), to $a$ evolving instantaneously as $a_\infty(V)$ (the reduced model), is that excitation must be much stronger in the reduced model to observe subtractive inhibition.  Typical values of $g_{Syn,E}$ in the full model are around 0.5 (see Fig.~\ref{fig:DivSubBoundary1}), and typical values of $g_{Syn,E}$ in the reduced model are around 3.  
This suggests that the speed of A-current activation (not just the strength of the A-current) plays a role in switching the effect of inhibition from divisive to subtractive.

\begin{figure}[!h]
\centering
\includegraphics[width=\textwidth]{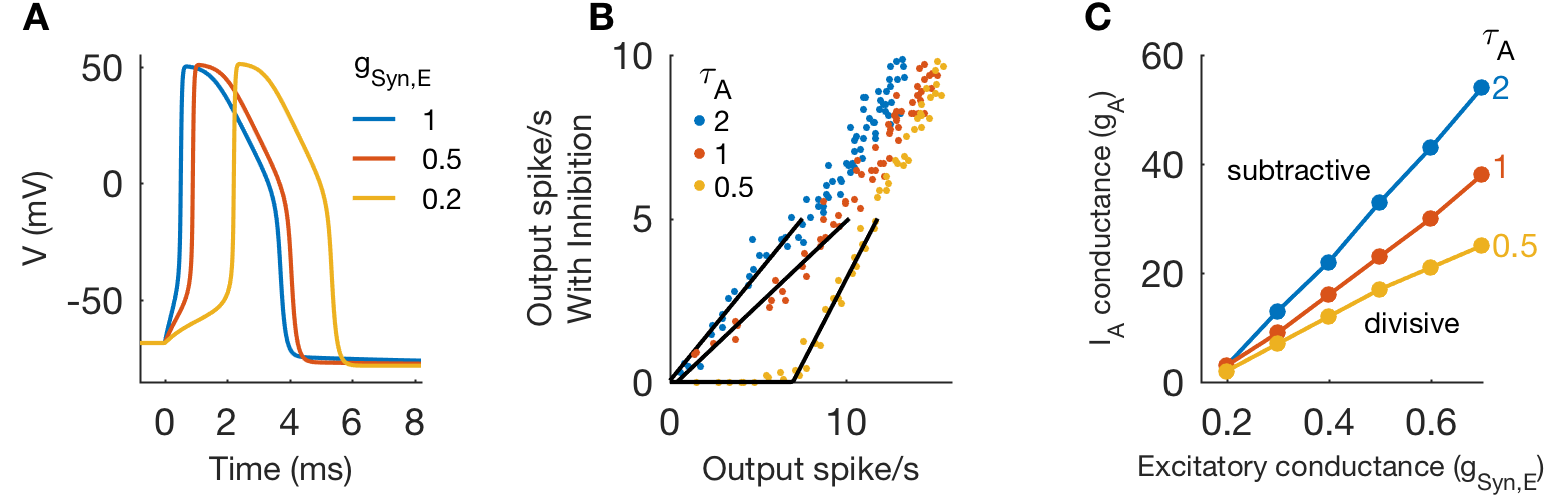}
\caption{Role of $\tau_A$ in determining switch between divisive and subtractive inhibition.
{\bf A:}  Voltage traces in response to excitatory inputs of varying strengths ($g_A=0$, and $g_{Syn,I}=0$).  
{\bf B:} Threshold-linear relation between output firing rates in simulations of one-compartment model with and without inhibition for varying A-channel activation time constant ($\tau_A=0.5, 1, 2$) and $g_A=20$. Synaptic parameter values are $g_{Syn,E}=0.5$, and (for simulations with inhibition) $g_{Syn,I}=1$ and $r_I=50$. 
{\bf C:}  Critical values of $g_A$ that define boundary between subtractive and divisive inhibition in $(g_{Syn,E}, g_A)$ parameter space. 
The boundary shifts downward as $\tau_A$ decreases, indicating that faster activating A-current enables inhibition to have a subtractive effect for lower values of $g_A$.
}
\label{fig:DivSubTauA}
\end{figure}

For inhibition to have a subtractive effect, responses to infrequent excitatory inputs must be suppressed.  In the reduced model, this occurs when $g_A$ is sufficiently strong because the A-type channel activates ``instantaneously'' and can prevent spike initiation.  In the one-compartment model with ``non-instantaneous'' $a$ variable, large $g_A$ could switch the effect of inhibition to subtractive, but only if excitatory input strength was also sufficiently small (recall Fig~\ref{fig:DivSubBoundary1}).
The importance of small $g_{Syn,E}$ is demonstrated in Fig~\ref{fig:DivSubTauA}. We show time-courses of voltage in the one-compartment model for $g_{Syn,E}=0.2, 0.5,\mbox{ and }1$, and with $g_A=0$ and $g_I=0$.  In all cases, the input evokes an output spike.  Notice, however, that as input strength weakens, there is a marked delay in the time before spike initiation.  For the weakest input used ($g_{syn,E}=0.2$), there is a delay of roughly 2~ms before the rapid upstroke of $V$ at the onset of the action potential.  During this ``pause,'' voltage is slowly ramping up and, simultaneously, recruiting additional A-current as the $a$ variable activates.  The amount of $I_A$ available to suppress spike initiation depends, therefore, on A-channel maximal conductance ($g_A$) and also the time-constant of $I_A$ activation ($\tau_A$).

From this observation, we draw the following conclusion: ``non-instantaneous'' $I_A$ can act to switch the effect of inhibition from divisive to subtractive, but only if it activates rapidly enough relative to the dynamics of spike initiation.  To illustrate our point, we simulated the model with three values of A-channel activation time constant ($\tau_A = 0.5, 1, 2)$, using $g_{Syn,E}=0.5, g_{Syn,I}=1,\mbox{ and } r_I=50~\mbox{Hz}$.  As shown in Fig~\ref{fig:DivSubTauA}B, inhibition is divisive for slower activation kinetics ($\tau_A=1,  2$) and subtractive for faster activation kinetics ($\tau_A=0.5$).  In Fig~\ref{fig:DivSubTauA}C, we map the boundary between divisive and subtractive inhibition in the $(g_{Syn,E}, g_A)$ parameter plane.  There is a strong effect of $\tau_A$.
Faster activation kinetics (smaller $\tau_A$ values) shift the critical point at which inhibition switches from divisive to subtractive to lower values values of $g_A$.

\subsection*{Results for multi-compartment model}

Our prior observation, that delaying spike initiation allows inhibition to have a subtractive effect for ``non-instantaneous'' A-channel activation, led us to investigate other cellular mechanisms that could have a similar effect. 
To this end, we considered a multi-compartment neuron model that describes a soma and passive dendrite.  Inhibition and voltage-gated currents are restricted to the soma compartment, and excitation targets a location somewhere on the dendrite.
Passive cable theory tells us that the amplitudes of excitatory post-synaptic potentials attenuate and their rising slopes become less steep as signals spread along the cable~\cite{Rinzel1974}. 
By varying the location of excitatory synaptic inputs to the dendrite in the multi-compartment model, we can, therefore, adjust the shape of excitatory post-synaptic potentials as they arrive in the soma.

\begin{figure}[!h]
\centering
\includegraphics[width=\textwidth]{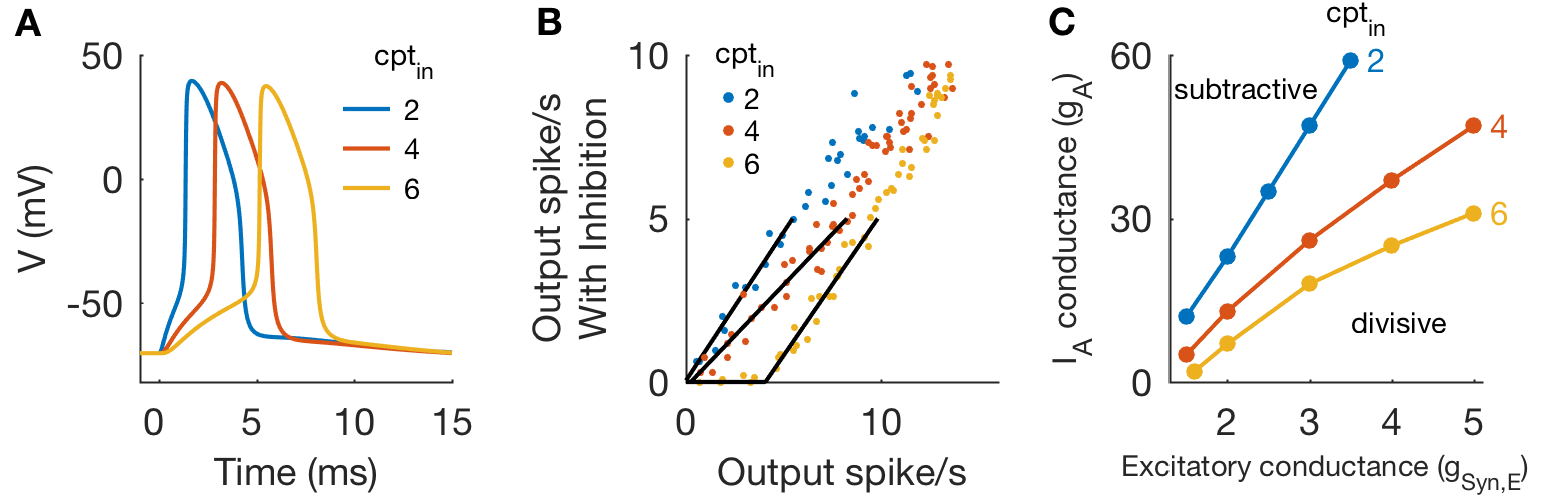}
\caption{Divisive and subtractive inhibition in a multi-compartment neuron model.
{\bf A:}  Voltage traces in response to excitatory inputs at varying input locations along the dendrite.
Parameter values in these simulations: $g_{Syn,E}=3$, $g_{Syn,I}=0$, and $g_A=0$.  
Inputs distant from the soma lead to spike initiation with millisecond-scale delay between excitatory input and spike onset.
{\bf B:} Threshold-linear relation between output firing rates in simulations of the multi-compartment model with and without inhibition for varying input location and $g_A=20$. For simulations with inhibition: $g_{Syn,I}=1$ and $r_I=50$.  Inhibition is subtractive for distal excitatory input ($cpt_{in}=6$).
{\bf C:}  Critical values of $g_A$ that define boundary between subtractive and divisive inhibition in $(g_{Syn,E}, g_A)$ parameter space.  
The boundary shifts downward as excitatory inputs are moved to more distal locations, indicating that inhibition has a subtractive effect for lower values of $g_A$ for more distal inputs.
}
\label{fig:DivSubMulti}
\end{figure}

Examples of action potentials, recorded in the soma compartment, in
response to inputs at different locations along the dendrite are shown
in Fig~\ref{fig:DivSubMulti}A.
Synaptic conductance strength is constant ($g_{Syn,E}=2$ in these simulations).  
Inputs that arrive proximal to the soma are large and fast-rising relative to responses to more distal inputs, and thus evoke action potentials with shorter latencies. 
The parameter $cpt_{in}$ identifies the compartment that receives synaptic excitation.  It takes values from 1 (proximal) to 9 (distal).

We included synaptic inhibition in the model (targeting the soma), and used simulations to characterize the effect of inhibition as either subtractive or divisive.  
In Fig~\ref{fig:DivSubMulti}B 
we observe a transition from divisive to subtractive inhibition as we move the location on the dendrite at which synaptic excitation targets the cell.  
This matches our expectation: synaptic excitation placed at more distant locations will generate weaker and slower rising inputs in the soma.  This will, in turn, lead to spikes that initiate slowly and that give time for A-channel conductance to activate and prevent spike generation.
   
We explored the $(g_A,\, g_{Syn,E})$ parameter plane and identified the boundary
separating regions in which inhibition has a divisive effect on firing
rates and regions in which inhibition has a subtractive effect
(following the procedure used previously for Fig~\ref{fig:DivSubBoundary1}).
We find that there is a dramatic effect of input location, as shown in Fig~\ref{fig:DivSubMulti}C.
The region of the $(g_A,\, g_{Syn,E})$ parameter plane over which
inhibition has a divisive effect is smaller when inputs are more distant from the soma.

\section*{Discussion}

Neurons process and convey information in the brain by converting
barrages of synaptic inputs into spiking outputs.  This transfer from
inputs to outputs is a highly complex process due to the inherently noisy and nonlinear nature of synaptic and neural processes.  Using a combination of computer simulation and mathematical analysis of biophysically-based neuron models, we have probed the relation between synaptic inputs and spiking outputs.  We found that the A-type potassium current (a fast-activating, negative feedback current) can act to switch the effect of inhibition on output firing from divisive to subtractive.  This provides a clear demonstration of how the internal dynamics of a neuron can control the functional impact of inhibition.  

\subsection*{Analysis identifies when  $I_A$ promotes divisive or subtractive inhibition}
Using simulations and phase plane analysis, we systematically investigated  conditions under which inhibition acts on firing rate outputs in a divisive or subtractive manner.  
We first identified critical values of $I_A$ conductance ($g_A$) at which the effect of inhibition switched from divisive (for lower $g_A$ values) to subtractive (for higher $g_A$ values) (Fig~\ref{fig:DivSubBoundary1}).  In the reduced model, we approximated this critical value of $g_A$ using bifurcation analysis.  By tracking the left-knee of the $V$-nullcline (Fig~\ref{knees}), we identified this critical value of $g_A$ as a bifurcation point at which the neuron model ceased to be excitable in response to synaptic inputs (Fig~\ref{leftknee}).  Key in this analysis was the separation of time scales between fast variables ($V$) and slow variables ($n$, $b$).  In fact, the inactivation variable $b$ was sufficiently slow that it could be treated as a constant with a value that depended on the input rate (Fig~\ref{bav}).  This simplification enabled further analysis. By viewing the spiking output of the model as a Poisson process modified by a refractory period and inhibition-dependent firing threshold, we approximated firing rates at the point of spiking onset (Fig~\ref{slope}) as well as for arbitrary input rates (Fig~\ref{fig:firingThreshold}). 

A-type potassium current is a source of dynamic, voltage-gated negative feedback that is fast activating.  
We leveraged this property to obtain analytical results by assuming
that the gating variable for $I_A$ activation, $a$, evolved instantaneously to its voltage-dependent equilibrium value (see also~\cite{Rush1995}).  
We also performed simulations without this assumption and discovered a delicate interaction between the speeds of $I_A$ activation and spike initiation.
In particular, subtractive inhibition required that $I_A$ is sufficiently strong and that it activates sufficiently rapidly to prevent spike initiation~(Fig~\ref{fig:DivSubTauA}).
For our standard value of $I_A$ activation ($\tau_A=2~\mbox{ms}$), we
found that, in conditions of slow spike initiation, $I_A$ could ``ramp
up'' during slowly-developing spikes and suppress spike initiation.
Weak excitatory inputs, or excitatory inputs targeting more
distal regions in a model that included a spatially-extended dendritic process (Fig~\ref{fig:DivSubMulti}) produced spikes that were slow to initiate, and were therefore scenarios in which inhibition was subtractive for low to modest levels of $g_A$. 

\subsection*{Relation to previous works}

Divisive inhibition is a mechanism of neural gain control and has been the subject of numerous studies; see~\cite{Silver2010} for review.  
We found that, at lower levels of $g_A$, inhibition can have a divisive effect on the input/ouput properties of a spiking neuron responding to a mixture of random excitatory inputs and periodic inhibitory inputs.  The amount of $g_A$ altered the slope (gain) of the output firing rate, and thereby tunes the gain control in this system.  This result is consistent with the results of a recent {\it in vitro} study of neurons in the rostral nucleus of the solitary tract ~\cite{Chen2016}.  In that experiment, Chen and colleagues controlled inhibition using optogenetic techniques and constructed threshold linear functions to express the relation between firing responses with and without inhibition (analogous to our Fig~\ref{fig:DivSub1}C, and similar figures).  They observed that slopes of threshold-linear function were more shallow for neurons with $I_A$, as compared to neurons in the same nucleus that did not have $I_A$.  Thus, the presence of $I_A$ enhanced the divisive effect of inhibition. 
Previous modeling work has identified similar gain control effects by $I_A$~\cite{Patel2012}.  

We observed, additionally, that at higher levels of $g_A$, the A-type current can switch the effect of inhibition from divisive to subtractive.  This demonstrates a novel example of how the internal dynamics of a neuron interact with synaptic inhibition to change the neuron's computational properties (input/output relation).  Previous studies have explored the multi-faceted ways in which $I_A$ current can alter neural dynamics.  Connor and Stevens established $I_A$ current as a mechanism to prolong interspike intervals of repetitively-firing neurons to arbitrary lengths (``type I'' firing dynamics)~\cite{Connor1971}.   Other identified functions of $I_A$ include prolonging first spike latency~\cite{Gerber1993}, producing burst firing patterns and preventing anodal break (rebound) firing~\cite{Rush1995}, filtering synaptic inputs in favor of slow time-scale NMDA receptor-mediated inputs~\cite{Schoppa1999}, and affecting the correlation in spiking among neurons responding to common inputs~\cite{Barreiro2012}.
Our contribution adds to the rich repertoire of $I_A$ function. 

\subsection*{Implications for modulation of inhibitory effects}

We have identified routes to subtractive inhibition that depends only on mechanisms that could be readily adjusted by processes of plasticity and neuromodulation.  
In particular, we have shown that strong and fast $I_A$ can lead to subtractive inhibition.  The strength and kinetics of the A-type Potassium channels can be modulated in a variety of ways~\cite{Birnbaum2004,An2006,Cai2007,Carrasquillo2014}.  For example, in neurons involved in gastro-intestinal function, A-type potassium channels were modified both by diet~\cite{Boxwell2015,Browning2013,Li2013} and gastric disorders~\cite{Li2014}.

We also found that weak excitatory inputs or more distally-located excitatory inputs led to subtractive inhibition by slowing the onset of action potentials.  
Synaptic plasticity and modulation of the electrical properties of dendrites can adjust the strength and propagation of excitatory inputs~\cite{Sjostrom2008}, and plasticity of spike initiation zones could change the dynamics of spike initiation~\cite{Adachi2015,Brette2013}.
These changes happen at the level of the output neuron.
They do not require ``global'' modulatory effects to change background network activity, circuit structure, or the balance of excitation and inhibition.  
We conclude, then, that $I_A$ can add flexibility to neural systems by allowing neurons to ``self-regulate'' whether inhibition acts in a subtractive or divisive manner.

\section*{Acknowledgments}
This work was partially supported by the NSF award DMS1410935  to DT and the NIH award R01DC016112 to JBT and DT. The authors thank Xueying Wang for helpful discussions and insights.

\section*{Conflict of Interest}
The authors declare no competing financial interests.

%\bibliography{bib}

\end{document}